\documentclass[aps,prb, preprint, longbibliography]{revtex4-1}

\usepackage{graphicx}
\usepackage{amssymb}
\usepackage{epstopdf}
\usepackage{dcolumn}
\usepackage{bm}
\usepackage{amsmath}
\usepackage{mathrsfs}
\usepackage{siunitx}
\usepackage{subfigure}
\usepackage{flafter}
\usepackage{caption}
\usepackage{tabularx}
\usepackage{float}
\usepackage{braket,mleftright}
\usepackage[font={small}]{caption}
\newcolumntype{C}[1]{>{\centering\let\newline\\\arraybackslash\hspace{0pt}}m{#1}}

\newcommand{\ep}{\varepsilon}
\newcommand{\kv}{{\bf k}}

\begin{document}

\title{Graphene Contacts to a HfSe$_2$/SnS$_2$ Heterostructure } 

\author{Shanshan Su}
\email{ssu008@ucr.edu}
\affiliation{Department of Electrical and Computer Engineering, University of California, Riverside, CA 92521, USA}

\author{Protik Das}
\affiliation{Department of Electrical and Computer Engineering, University of California, Riverside, CA 92521, USA}

\author{Supeng Ge}
\affiliation{Department of Physics and Astronomy, University of California, Riverside, CA 92521, USA}

\author{Roger K. Lake}
\email{rlake@ece.ucr.edu}
\affiliation{Department of Electrical and Computer Engineering, University of California, Riverside, CA 92521, USA}
\date{\today}

\begin{abstract}
Placing graphene on SnS$_2$ results in significant charge transfer, 
on the order of $10^{13}/{\rm cm}^2$, 
from the graphene to the SnS$_2$, 
and the charge transfer results in a negative Schottky barrier contact for 
electron injection from the graphene into the SnS$_2$ conduction band. 
However, due to the $s-p_{x,y}$ composition of the SnS$_2$ conduction band, 
the coupling between the SnS$_2$ and the graphene is relatively weak. 
A third layer, HfSe$_2$, placed between the SnS$_2$ and the graphene, 
serves as a ‘matrix element matching layer,’ since it has strong coupling to both 
the graphene and the SnS$_2$. 
It increases the coupling to the graphene by a factor of 10, 
and it has little effect on the negative Schottky barrier height,
since the conduction band wavefucntion of the SnS$_2$ / HfSe$_2$ is a 
coherent superposition of the orbitals from the two individual 
layers, such that there is no energy barrier for an electron to move between the two layers.
This paper first investigates the electronic properties of the 
heterostructure bilayer SnS$_2$ / HfSe$_2$ in the presence of an applied vertical electric field, 
and then it investigates the trilayer systems of BN / SnS$_2$ / HfSe$_2$ 
and graphene / SnS$_2$ / HfSe$_2$. 
A tunneling Hamiltonian estimate of the the contact resistance of the graphene to the 
SnS$_2$ / HfSe$_2$ heterostructure indicates an excellent low-resistance contact. 
\end{abstract}

\maketitle 

\section{Introduction}

Heterostructures of two-dimensional (2D) 
van der Waals (vdW) materials are being extensively investigated
\cite{KJ_Band_Alignment_APL13, J_Wu_Band_Alignment_APL13, fu_direct_2016, lee2013flexible, lee_atomically_2014, balu_effect_2012, su2016modulating}.
Recent studies of vdW heterostructures have shown that it is possible to build type II heterojunctions 
and nearly broken gap 
heterojunctions \cite{huang_strain_2015,huang_electric-field_2015,kosmider_electronic_2013, bernardi2013extraordinary, li_two-dimensional_2015, li_two-dimensional_2015, schlaf1999band}.
In type II heterojunctions, the electron-hole pair is separated both spatially and energetically 
enabling efficient photovoltaics and photodetection \cite{kosmider_electronic_2013, bernardi2013extraordinary}.
A few heterostructures composed of 2H transition metal dichalcogenides, such as WSe$_2$/MoSe$_2$, 
remain direct gap with the conduction and valence bands at $K$.
The majority of heterostructures, such as, for example,
black phosphorus/MoS$_2$ \cite{huang_strain_2015,huang_electric-field_2015}
are indirect gap, with, in this particular case, the valence band at $\Gamma$ and the conduction
band at $K$.
For electronic applications,
multi-layer stacks of 2D materials such as 
black phosphorus/SnSe$_2$ \cite{yan2015esaki}, WSe$_2$/SnSe$_2$ \cite{li_two-dimensional_2015,schlaf1999band}, 
graphene/BN/graphene \cite{britnell2012field} and graphene/WS$_2$ \cite{georgiou2013vertical}
are being exploited for tunnel devices and tunneling field-effect transistors (TFETs).
There is also interest in using graphene to create direct bandgaps in multilayer 
heterostructures \cite{terrones2013novel, ghorbani2016effect, hu2016effects}, 
using graphene to make contact or tune other 2D materials 
\cite{Graphene_MoS2_Hybrid_Palacios_NLett14,pandey2016pressure,jin2015tuning,shi2015all,ebnonnasir2014tunable,liu2015electric,Graphene_WSe2_LEAST_NScale16,XDuan_Graphene_MoS2_NLett15,Neg_Schottky_Graphene_MoS2_Qiu_SRep15},
and to tune the workfunction to enhance cold cathode emission \cite{Graphene_SnS2_APL14}.

In the last application \cite{Graphene_SnS2_APL14}, 
placing graphene on SnS$_2$ significantly reduced the workfunction
from that of SnS$_2$ alone, and the charge transfer between the two materials resulted in p-type graphene
and n-type SnS$_2$.
The Fermi level of the composite aligned above the conduction band minimum of the SnS$_2$.
From an electrical contact point of view, such an energetic alignment is a negative Schottky barrier contact,
and it is highly desirable, since it gives a low contact resistance.
But there are two barriers to inter-layer current flow. 
One barrier is the energetic barrier represented by the Schottky barrier height.
In the graphene / SnS$_2$ system, this barrier is negative, so it is very favorable.
The other barrier is the inter-layer coupling between the two layers.
This coupling depends on the matrix element between the Bloch functions
of the bands in each layer that are being coupled. 
This matrix element will depend on the orbital composition of the bands and their 
positions in $k$-space \cite{KZhou_rotated_MoS2}. 
We find that this coupling is weak between graphene and SnS$_2$ near the Fermi level.
To improve this coupling while maintaining a negative Schottky barrier,
we investigate the use of a third material, HfSe$_2$, that serves as a 
`matrix element matching' layer between the SnS$_2$ and the graphene, since it has good 
coupling to both the SnS$_2$ and the graphene. 
%

In this paper, we analyze a multi-layer structure composed of 
monolayer HfSe$_2$, SnS$_2$, graphene, and BN.
Both HfSe$_2$ and SnS$_2$ are 1T polytype, hexagonal, 2D materials with indirect band gaps.
In monolayer form, their conduction bands are at $M$, and their
valence bands are at $\Gamma$.
Stacking the two layers together creates an indirect-gap heterojunction
that has type II qualities, but it does not fall cleanly into any one of the categories
used to classify heterostructures of three-dimensional semiconductors,
i.e. type I, type II, or type III,
since the conduction bands strongly couple, and the wavefunction is distributed across both layers.
An electric field applied to the heterostructure causes a shift in
weight of the conduction band wavefunction from the HfSe$_2$ layer to the SnS$_2$ layer
such that the band alignment takes on a type I quality.
A commensurate stacking on graphene or BN using a $2\times 2$ supercell of
the HfSe$_2$ $/$ SnS$_2$ and a $3\times 3$ supercell of the graphene or BN 
zone-folds the $M$ point of the  HfSe$_2$ $/$ SnS$_2$ back to
$\Gamma$, and it zone-folds the $K$ point of the graphene or BN back to $\Gamma$
resulting in a direct-bandgap heterostructure.
The strain between the two systems is low, 1.7\% for the BN and 0.1\% for the graphene.
The charge transfer from the graphene to the HfSe$_2$ $/$ SnS$_2$ results in a negative Schottky
barrier contact to the conduction band.

This paper is organized as follows. 
Sec. \ref{sec:Method} describes the methods 
based on density functional theory. 
In Sec. \ref{sec:results},
AA and AB stacked heterostructures of
HfSe$_2$ $/$ SnS$_2$ are first analyzed, and the effect of an applied
vertical electric field is described. 
Then a third layer of either graphene or BN is added, 
and the tri-layer structures are analyzed and discussed.
Conclusions are presented in Sec. \ref{sec:conclusion}.

\section{Method}
\label{sec:Method}

Density functional theory calculations 
are performed with the Vienna  {\it ab initio} simulation package (VASP)\cite
{kresse_efficient_1996,kresse_ab_1993,kresse_efficiency_1996} in the projected-augmented-wave method \cite
{blochl_projector_1994}. 
The generalized gradient 
approximation (GGA) of the Perdew-Burke-Ernzerhof form \cite{perdew_atoms_1992, wang_correlation_1991, 
kresse_ultrasoft_1999} 
is used for the exchange correlation energy. 
The vdW interactions are included with the DFT-D2 method of Grimme \cite{harl_assessing_2010}. 
The kinetic energy cutoff is 500 eV for all calculations.
The first Brillouin zone is sampled with a 8 $\times$ 8 $\times$ 1 $\Gamma$-centered Monkhorst-Pack grid. 
During all structural relaxations, the convergence tolerance on the 
Hellmann-Feynman forces is less than 0.01 eV\AA . 
A vacuum layer larger than 25 ${\rm \AA}$ is used for heterostructures to eliminate the interaction between adjacent images
in the vertical direction.
To determine more quantitative values for bandgaps, calculations
are also performed with the
hybrid Heyd-Scuseria-Ernzerhof (HSE) functional\cite{heyd_hybrid_2003}. 
The HSE calculations incorporate 25\% short-range Hartree-Fock exchange.
The screening parameter $\mu$ is set to 0.2 \AA $^{-1}$.

The optimized lattice constant of SnS$_2$ is 3.69 \AA , 
and the optimized lattice constant of HfSe$_2$ is 3.72 \AA .
The lattice mismatch between SnS$_2$ and HfSe$_2$ is less than 1\%.
The lattice constant of the heterostructure is set to the average value of 3.70 \AA .
Monolayer HfSe$_2$ and SnS$_2$ are with bandgaps of 1.1 eV\cite{gaiser2004band} and 2.4 eV\cite{koda2016coincidence}, respectively. 
As a check of the sensitivity of the electronic bandstructure
to the lattice constant, we considered the two extreme cases resulting from
exchanging the lattice constants of HfSe$_2$ and SnS$_2$, 
and re-calculating the bandstructures of the individual layers.
We found that the bandstructures of the individual material remained almost the same.

\section{results and discussion} 
\label{sec:results}

\begin{figure}
\includegraphics[width=.5\linewidth]{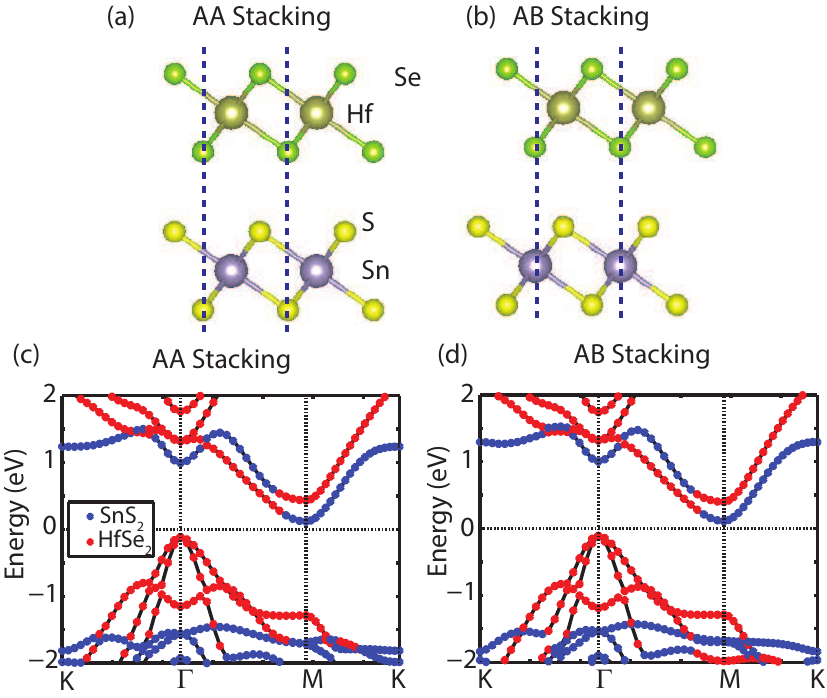}
\caption{\label{fig:fig1} 
Atomic structure of (a) AA stacking and (b) AB stacking. 
(c) AA electronic structure and (d) AB electronic structure.}
\end{figure}

As shown in Fig. \ref{fig:fig1}(a), 
we consider AA and AB stacking of HfSe$_2$ on SnS$_2$.
Both 1T bulk HfSe$_2$ and SnS$_2$ stack in AA order in which
the metal atoms of one layer align with the metal atoms of the other.
Sliding one layer with respect to the other such that the the metal atoms of
one layer align with the chalcogenide atoms of the other gives AB stacking.
AA and AB stacking correspond to the two most stable stacking geometries.
The total energy of the AA and AB stacked heterostructures are -6.053 eV per atom and -6.050 eV per atom, 
respectively.
In $k$-space, both stacking arrangements have an indirect band gap
with the conduction band minimum (CBM) at the $M$ point
and the valence band maximum (VBM) at $\Gamma$.
In real-space, 
the CBM is more heavily weighted on the SnS$_2$ layer, 
and the VBM is localized on the HfSe$_2$ layer. 
The colors of the electronic bands in Fig. \ref{fig:fig1}(c,d) indicate the layer on which the wavefunction is
most heavily weighted as indicated in the legend.
%
%
%
The PBE indirect bandgaps are 237 meV and 224 meV for the AA and AB structures, respectively. 
There are two conduction bands close to each other at the $M$ point, 
and they originate from the conduction bands of the two individual layers.
The energy gap between the two lowest conduction bands is 
316.5 meV and 285.3 meV for AA and AB stacking, respectively.

A more quantitative determination of the energy spacings is made by
calculating the electronic structure using the HSE hybrid functional.
Qualitatively, the orbital composition and the order of the bands remain the same
and the primary difference is that the conduction-valence bandgaps increase.
The bandgaps increase to 0.88 eV and 0.89 eV for the AA and AB structures, respectively. 
However, the energy separation between the two conduction bands at $M$ remains essentially unchanged
with energies of 
317.6 meV and 281.4 meV for AA and AB structures, respectively.
This energy spacing between the two conduction bands is the critical energy 
that governs the crossing of the two conduction bands
under an applied electric field.
Since both the HSE and PBE calculations predict the same energy separation,
we use the computationally less expensive PBE functional to predict the behavior of the
heterostructure under applied cross-plane (vertical) electric fields.
Furthermore, since the VBM remains strongly localized in the HfSe$_2$
for all electric fields and multi-layer structures, the focus of the 
rest of the paper will be on the two lowest conduction bands
and their evolution with electric field and in contact with graphene or BN.

\begin{figure}
\includegraphics[width=.5\linewidth]{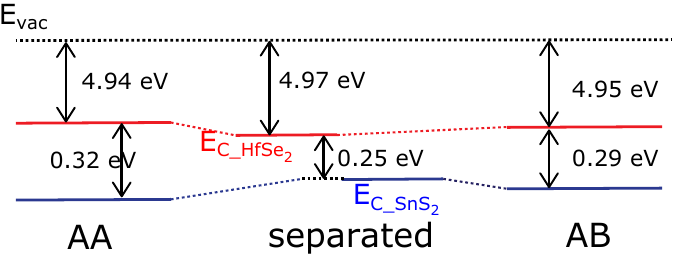}%
\caption{\label{fig:band_align_sep} 
The two lowest conduction band edges of the spatially separated HfSe$_2$/SnS$_2$ system (center), 
the AA stacked heterostructure (left), and AB stacked heterostructure (right). 
$E_{vac}$ is the energy level of vacuum.
In the well-separated case, 
the blue line is the conduction band edge of SnS$_2$, 
and the red line is the conduction band edge of HfSe$_2$.
For the AA and AB heterostructures, 
the red and blue lines indicate on which layer the conduction band edge wavefunction
is most heavily weighted.
}
\end{figure}

When the two monolayers are brought together,
the orbitals of the CBM in each layer will couple and push apart in energy.
To understand the evolution of the bands as the two materials are brought
together, we perform a DFT calculation of the AA structure with the two layers
separated by 2 nm.
This is sufficiently far apart that the bands do not interact,
but a common Fermi level is enforced giving the band lineup of the
well-separated, equilibrated, but non-interacting layers.
The conduction band alignment of the separated system is shown in Fig. \ref{fig:band_align_sep}.
When the layers are well-separated spatially, the energy separation of the
two conduction bands is 0.25 eV. 
When the two layers are brought together to form the heterostructure,
the two conduction bands push further apart by 40 meV for AB stacking and 70 meV for AA stacking.
This increase in energy separation is related to the coupling between the two
bands, and the larger splitting in the AA structure indicates
stronger coupling between the two conduction bands in that stacking arrangement.

For the spatially separated structure,
the two conduction bands are 100\% localized on the individual layers.
The lower conduction band is localized on the SnS$_2$, and the upper conduction band is
localized on the HfSe$_2$.
In the SnS$_2$, the conduction band wavefunction is weighted 54\% on the Sn, with 89\%
of that contribution from the s orbital, and 45\% on the S, with
83\% of that contribution from the $p_x$ and $p_y$ orbitals.
In HfSe$_2$, the conduction band wavefunction is weighted 79\% on the Hf. 
96\% of that comes from the d orbitals with the heaviest weight of 35\% coming
from d$_{z^2}$. 
The 21\% contribution from Se is 61\% from the p$_z$ orbital, 22\% from the d orbitals,
and 9\% from the s orbital.

When the two layers are brought together to form
the AA heterostructure, the magnitude squared of the 
CBM wavefunction no longer remains localized on the SnS$_2$,
but becomes distributed across both layers.
It is weighted approximately 60\% on the SnS$_2$ and 40\% on the HfSe$_2$. 
For the AB heterostructure, 
the wavefunction is weighted slightly more heavily on the SnS$_2$,
with a weight of 67\% on the SnS$_2$ and 33\% on the  HfSe$_2$.  
The fact that the CBM wavefunction is weighted more heavily on the SnS$_2$ layer
is consistent with the weaker coupling inferred from the smaller spitting of the bands
in the AB structure.
The orbital compositions of the individual layers remain qualitatively the same
as those of the isolated layers.
The VBM always remains localized in the 
HfSe$_2$ with an orbital composition from the $p_x$ and $p_y$ orbitals of the Se.

For both stacking arrangements, there is strong hybridization of the conduction band wavefunctions
of the two individual layers, 
and the conventional spatially resolved band picture
illustrated in Fig. \ref{fig:figheteAlign}(a) does not provide 
a good representation of the physics, at least for the conduction band.
Instead, for the conduction band, the picture of bonding and anti-bonding molecular orbitals is more
faithful to the underlying physics.
In this picture the lower and upper conduction band at the M point are the bonding and anti-bonding
combination of the isolated conduction bands of the individual layers.
As the two layers are brought together, the bands couple and push apart
resulting in two levels with the bonding orbital more heavily weighted on the SnS$_2$.
However, there is no energy barrier for an electron to move between the two layers,
since the CBM wavefunction is a coherent superposition of the orbitals on
both layers, and the probability of finding an electron on the SnS$_2$ layer is
at most a factor of 2 larger than on the HfSe$_2$ layer.

\begin{figure}
\includegraphics[width=.5\linewidth]{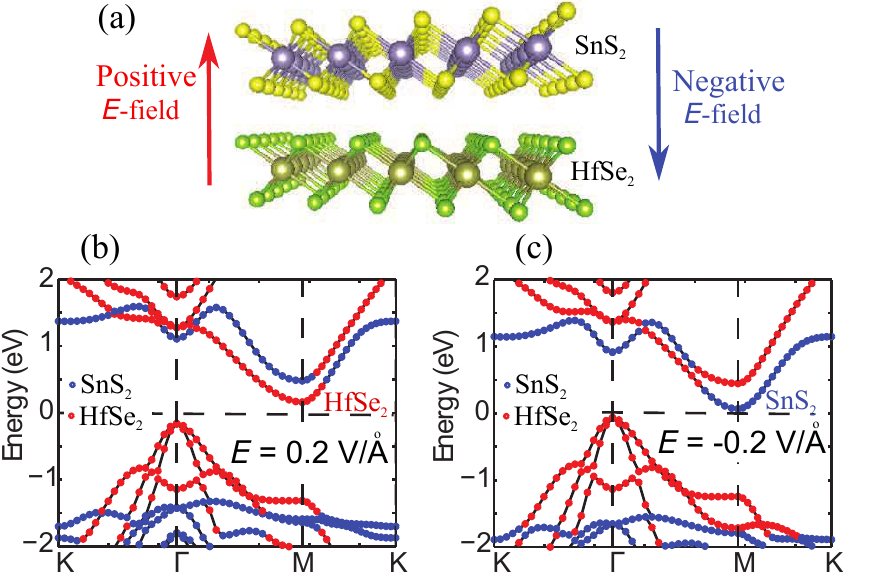}
\caption{\label{fig:field} 
(a) The AA heterostructure with arrows showing 
the direction of the applied electric fields (${\cal E}$) for the energy wavevector plots underneath.
(b) Bandstructure of the AA heterostructure under forward bias (${\cal E} = 0.2$ V/\AA ).
The HfSe$_2$ and SnS$_2$ CBMs have switched.
(c) Bandstructure of the AA heterostructure under reverse bias (${\cal E} = -0.2$ V/\AA ).
}
\end{figure}

\begin{figure}
\includegraphics[width=.5\linewidth]{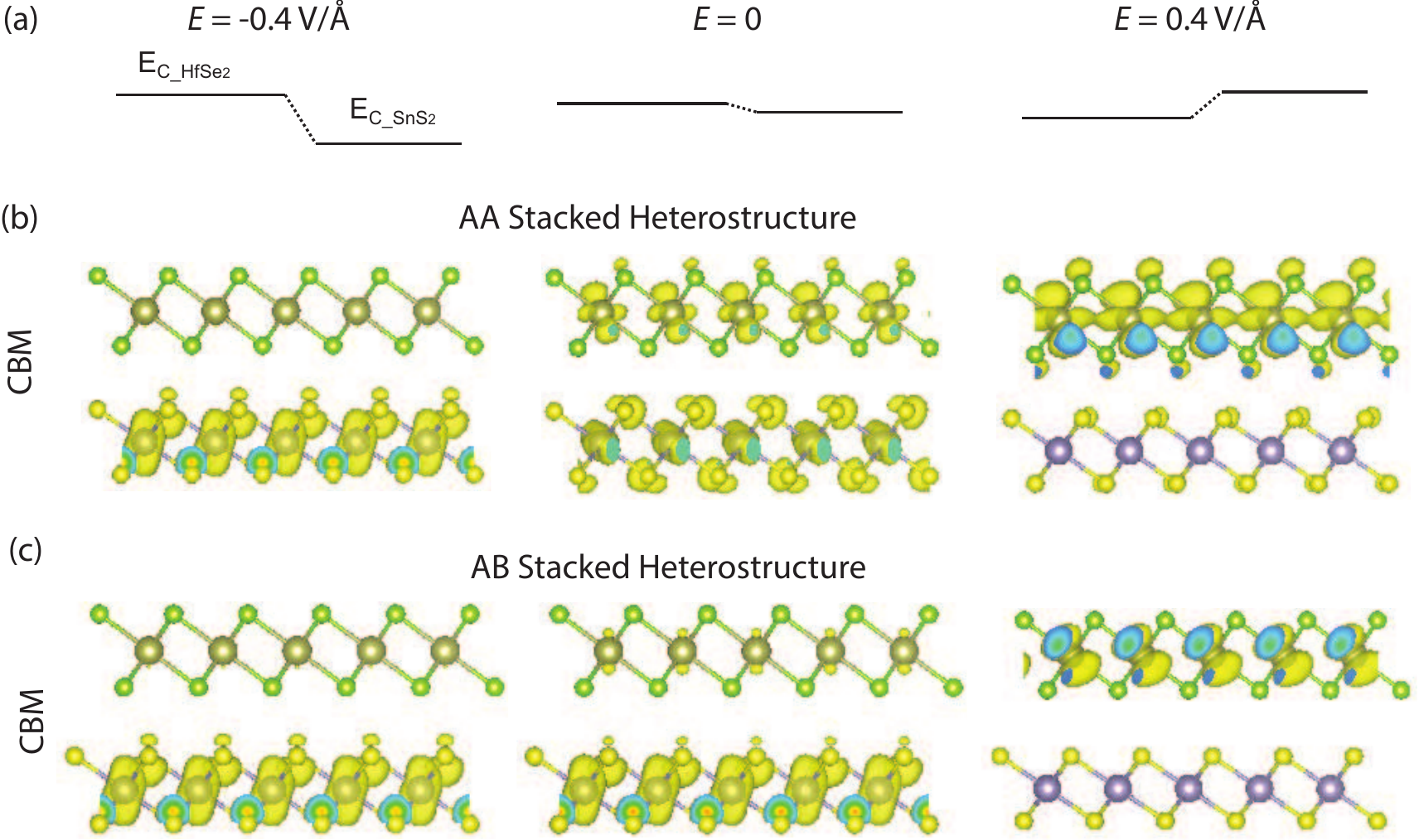}%
\caption{\label{fig:figheteAlign} 
(a) Schematic view of band alignments under -0.4 V/\AA , 0 V/\AA ,  and 0.4 V/\AA . 
(b) Isosurfaces of the orbital resolved wave-functions of the CBM of 
the AA stacked heterostructure under applied electric fields of -0.4, 0, and 0.4 V/\AA .
(c) Isosurfaces of band resolved wave-functions of the CBM andof the AB stacked heterostructure under 
applied electric fields of -0.4, 0, and 0.4 V/\AA .
}
\end{figure}

By applying an electric field, the relative weights 
on each layer of the first conduction band can be altered and even reversed.
As illustrated in Fig. \ref{fig:field}(a),
a positive electric field corresponds to a positive voltage applied to the HfSe$_2$ layer,
which means that the energy levels of the HfSe$_2$ layer are lowered with respect
to those in the SnS$_2$ layer.
Figs. \ref{fig:field}(b) and (c) show the AA electronic structure under forward and reverse bias,
respectively.
Under foward bias, the spectral weight of the CBM switches from the SnS$_2$ to the HfSe$_2$. 
The orbital compositions of the CBMs are illustrated in
Figs. \ref{fig:figheteAlign}
for three different electric fields, -0.4 V/\AA, 0 V/\AA, and 0.4 V/\AA,
corresponding to the left, middle, and right columns, respectively.
Fig. \ref{fig:figheteAlign}(a) illustrates the band alignments 
with the usual band picture used for bulk semiconductor heterojunctions
for the 3 different electric fields.
The left side of each band diagram represents the HfSe$_2$ layer and the right side
represents the SnS$_2$ layer.
The energy level of the CBM only indicates on which layer it is more heavily weighted.

\begin{figure}
\includegraphics[width=.5\linewidth]{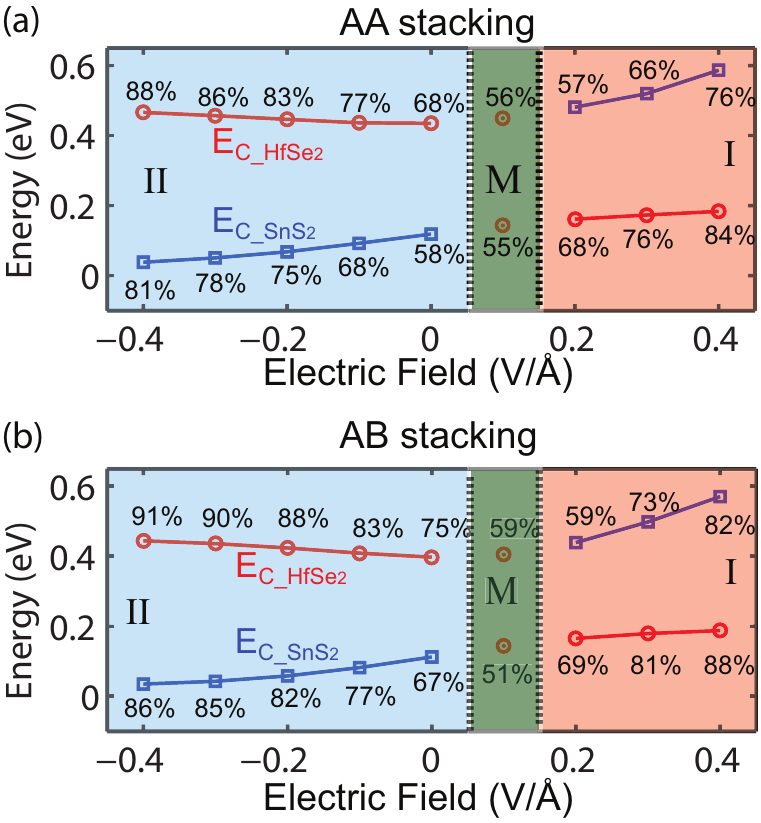}%
\caption{\label{fig:band_evolution} 
(a) Evolution of the two lowest conduction band edges as a function of electric field
for the (a) AA and (b) AB heterostructures. 
$E_{C_{\rm HfSe{_2}}}$ is the CBM of HfSe$_2$, 
and $E_{C_{\rm SnS{_2}}}$ is the CBM of SnS$_2$.
The weights of HfSe$_2$ and SnS$_2$ are marked on $E_{C_{\rm HfSe{_2}}}$ and $E_{C_{\rm SnS{_2}}}$, respectively.
Region II and region I correspond to heterojunction type II and type I respectively.
Region M represents the region that the two lowest conduction bands with heavily mixed contributions from the two layers of materials.
}
\end{figure}

As the electric field is ramped from negative to positive, the spectral weight
gradually shifts from the 
SnS$_2$ to the HfSe$_2$.
%
%
This shift of the wavefunction is illustrated in 
Fig. \ref{fig:band_evolution} for the AA and AB heterostructures.
Fig. \ref{fig:band_evolution} shows the energies of the two lowest CBMs
with the energy reference taken as the middle of the PBE bandgap.
The percentages give the percent spectral weight of the wavefunction
on the layer indicated by the label adjacent to each line.
The trends and quantitative values for
the AA and the AB stacked heterostructures are very similar.
At zero field, the CBM is weighted towards the SnS$_2$ as previously discussed.
At negative fields, the CBM wavefunction becomes more localized on the SnS$_2$ layer.
A shift in the CBM weight from the SnS$_2$ to the HfSe$_2$ layer occurs between
positive fields of 0.1 and 0.2 V/\AA . 
As the field becomes more positive the CBM wavefunction becomes more localized
on the HfSe$_2$.

Two dimensional materials will be in contact with other materials
as contacts, substrate, or encapsulation to prevent oxidation.
All-2D systems are very attractive since the interfaces are 
self-passivated and devoid of dangling bonds.
BN is a good insulator, and has recently been demonstrated to protect
highly reactive black phosphorous from oxidation \cite{Lau_BP_2DMat14}.
Graphene, a good conductor, is closely lattice matched to BN.
A $3\times3$ supercell of graphene or BN is also very closely lattice matched to a 
$2\times 2$ supercell of HfSe$_2$/SnS$_2$.
The lattice constants of BN and graphene are 2.51 \AA , and 2.47 \AA , respectively.
The lattice mismatches between the BN or graphene $3\times 3$ supercells and the 
HfSe$_2$/SnS$_2$ $2\times 2$ supercell are 1.7\% and 0.1\%, respectively. 
The lattice constant of the supercell is fixed to be the lattice constant of the 
HfSe$_2$/SnS$_2$ heterostructure, 
so that the the HfSe$_2$/SnS$_2$ heterostructure remains unstrained.

\begin{figure}
\includegraphics[width=3.5in]{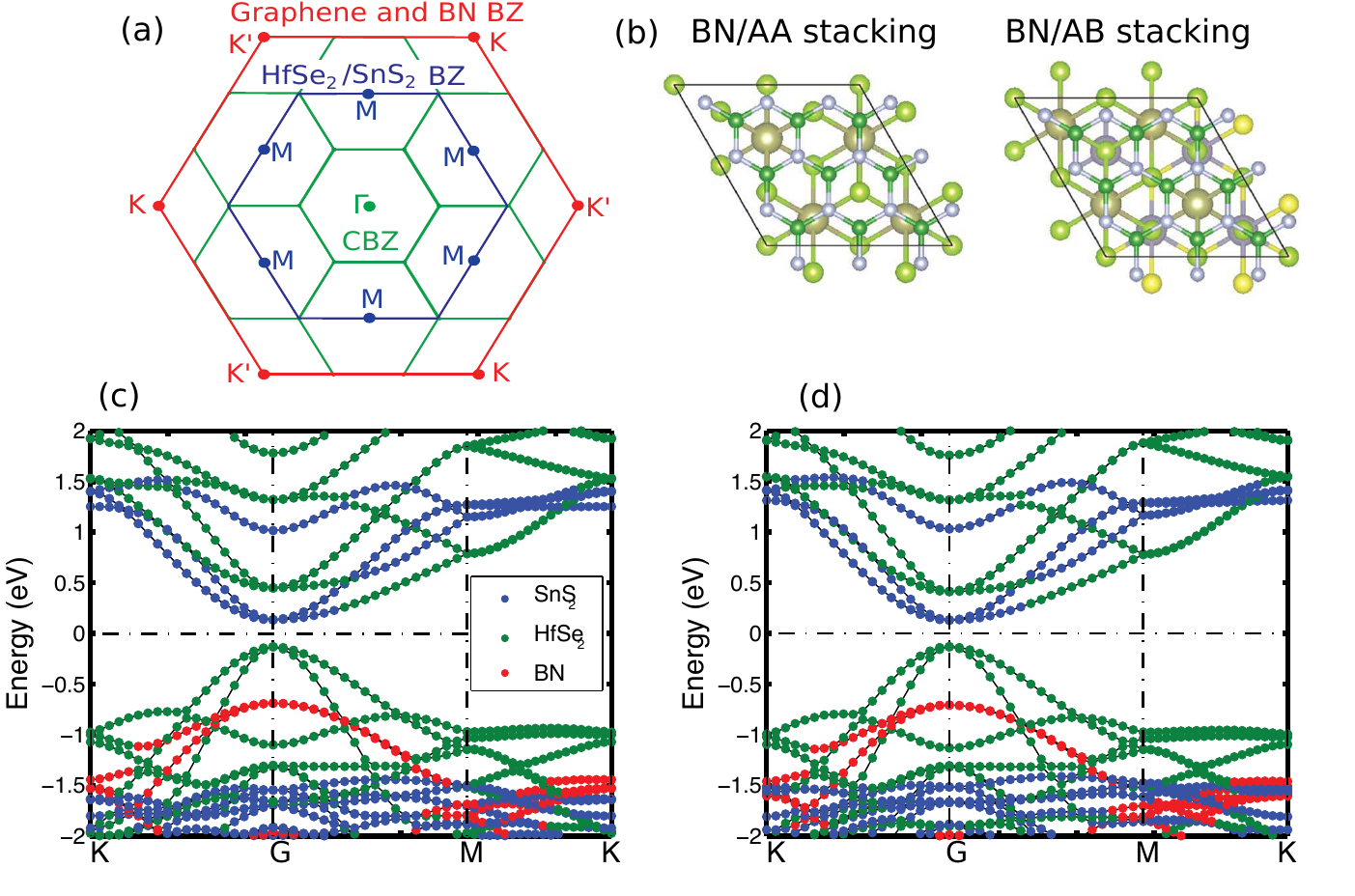}
\caption{\label{fig:trilayer} 
BN on HfSe$_2$/SnS$_2$. 
(a) Brillouin zone folding resulting from the $2\times 2$ unit cell of the 
HfSe$_2$/SnS$_2$ and the $3\times 3$ unit cell of the BN or graphene. 
The outer Brillouin zone (BZ) in red is the BZ of the graphene or BN. 
The BZ in blue is the BZ of the HfSe$_2$/SnS$_2$. 
The innermost BZ in green is the commensurate BZ (CBZ) of the supercell. 
The CBZ is tiled over the entire $k$-space region to show that the
$M$ point of the blue BZ lies at the $\Gamma$ point of the first repeated CBZ,
and the $K$ point of the red BZ lies at the $\Gamma$ point of the second repeated CBZ. 
(b) Atomistic structures for BN on AA and AB stacked HfSe$_2$/SnS$_2$. 
(c) Electronic bandstructure for AA stacking. 
(d) Electronic bandstructure for AB stacking. 
In (c) and (d), the color indicates on which layer the wavefunction is most heavily weighted.
}
\end{figure}

\begin{figure}
\includegraphics[width=.5\linewidth]{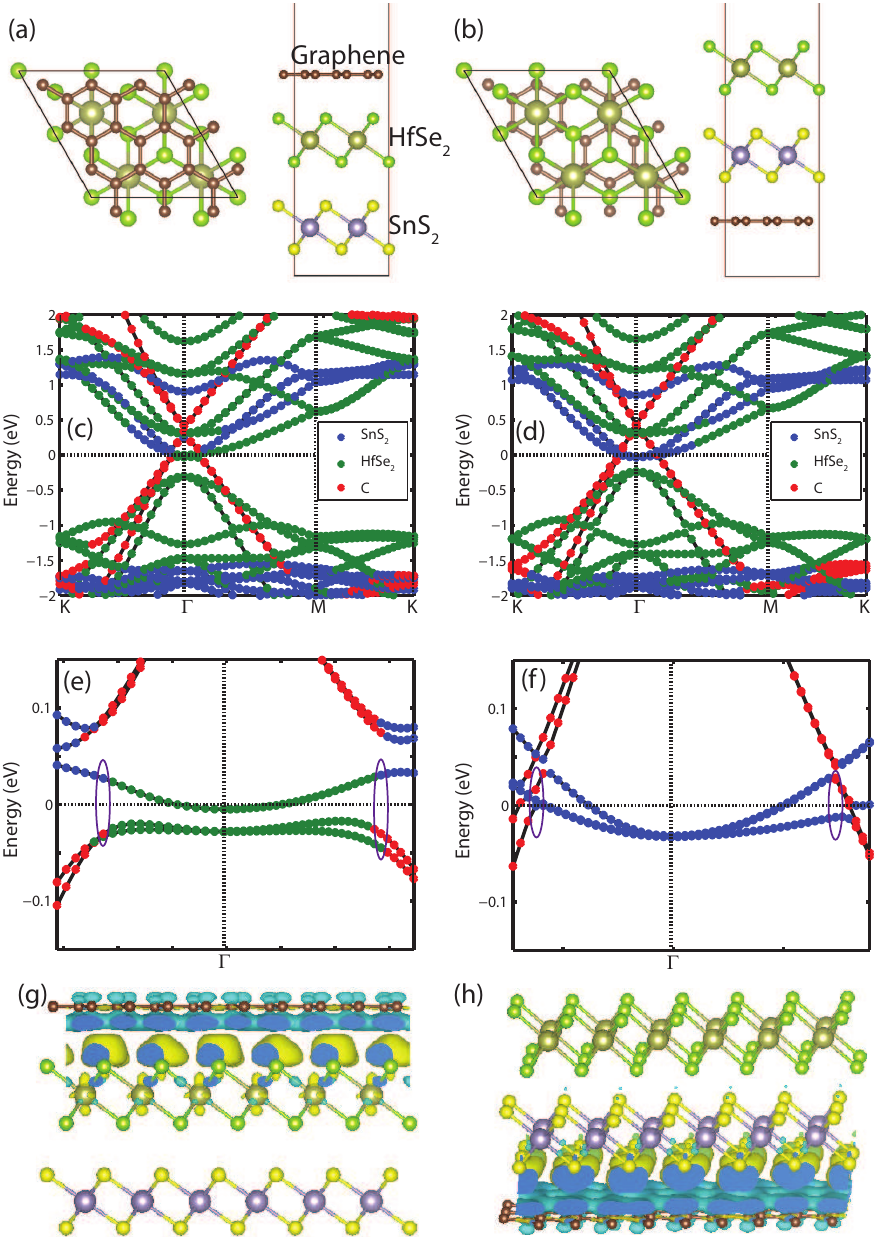}
\caption{\label{fig:AA_graphene_trilayer} 
Trilayer of graphene and AA stacked HfSe$_2$/SnS$_2$.
(a) Atomic structure of graphene on the HfSe$_2$ layer. 
(c) and (e) are the corresponding electronic structure plots. 
(e) focuses on the small energy range near the Fermi level.
(g) shows the charge transfer at the interface.
(b) Atomic structure of graphene on the SnS$_2$ layer.
(d) and (f) are the corresponding electronic structure plots. 
(f) focuses on the small energy range near the Fermi level.
(h) shows the charge transfer at the interface.
In (g) and (h), the charge accumulation and depletion is denoted by the yellow and blue color, respectively. 
The Fermi level is at $E=0$.
The purple circles in (e) and (f) indicate the anti-crossing of the
graphene hole band and the HfSe$_2$/SnS$_2$ conduction band.
}
\end{figure}

The tri-layer systems are stable. 
The formation energies are negative, and they are listed below in units of 
eV per primitive unit cell (UC) of the BN or graphene, where one primitive unit cell
consists of two atoms.
For BN on HfSe$_2$/SnS$_2$ as shown in Fig. \ref{fig:trilayer}, the formation energy is -0.087 eV/UC.
The formation energies for graphene on the HfSe$_2$ side or the SnS$_2$ side of the 
AA stacked heterostructure shown in Fig. \ref{fig:AA_graphene_trilayer}
are $-0.232$ eV/UC or $-0.185$ eV/UC, respectively.
The formation energies for graphene on the HfSe$_2$ side or the SnS$_2$ side of the 
AB stacked heterostructure are very similar, and they are
$-0.240$ eV/UC and $-0.193$ eV/UC, respectively.
Therefore, the trilayer systems are stable, and the graphene tri-layer structures are most
stable.

Fig. \ref{fig:trilayer}(a) shows the structure and supercell of
a BN monolayer on the HfSe$_2$ layer of the HfSe$_2$/SnS$_2$ heterostructure.
The electronic bandstructures for the AA and AB heterostructures with BN on the HfSe$_2$ layer
are shown in Figs. \ref{fig:trilayer}(b) and (c), respectively.
The bands of the HfSe$_2$/SnS$_2$ layers show no noticable change due to the proximity
of the BN. 
The BN bands are far from the Fermi energy and are buried deep in the valence and conduction bands of the
HfSe$_2$/SnS$_2$ as one would expect for a wide bandgap insulator.
Only the BN valence band can be seen on this energy scale.
What is most notable about this energy-momentum plot is that all of the band edges
now occur at $\Gamma$.
This is a result of zone-folding.
The $2\times 2$ supercell of the HfSe$_2$/SnS$_2$ folds the $M$ points back to $\Gamma$,
and the $3\times 3$ supercell of the BN folds the $K$ points back to $\Gamma$.
Thus, the system becomes direct gap.
The zone-folding of the different Brillouin zones is illustrated in Fig. \ref{fig:trilayer}(a).

\begin{figure}
\includegraphics[width=.5\linewidth]{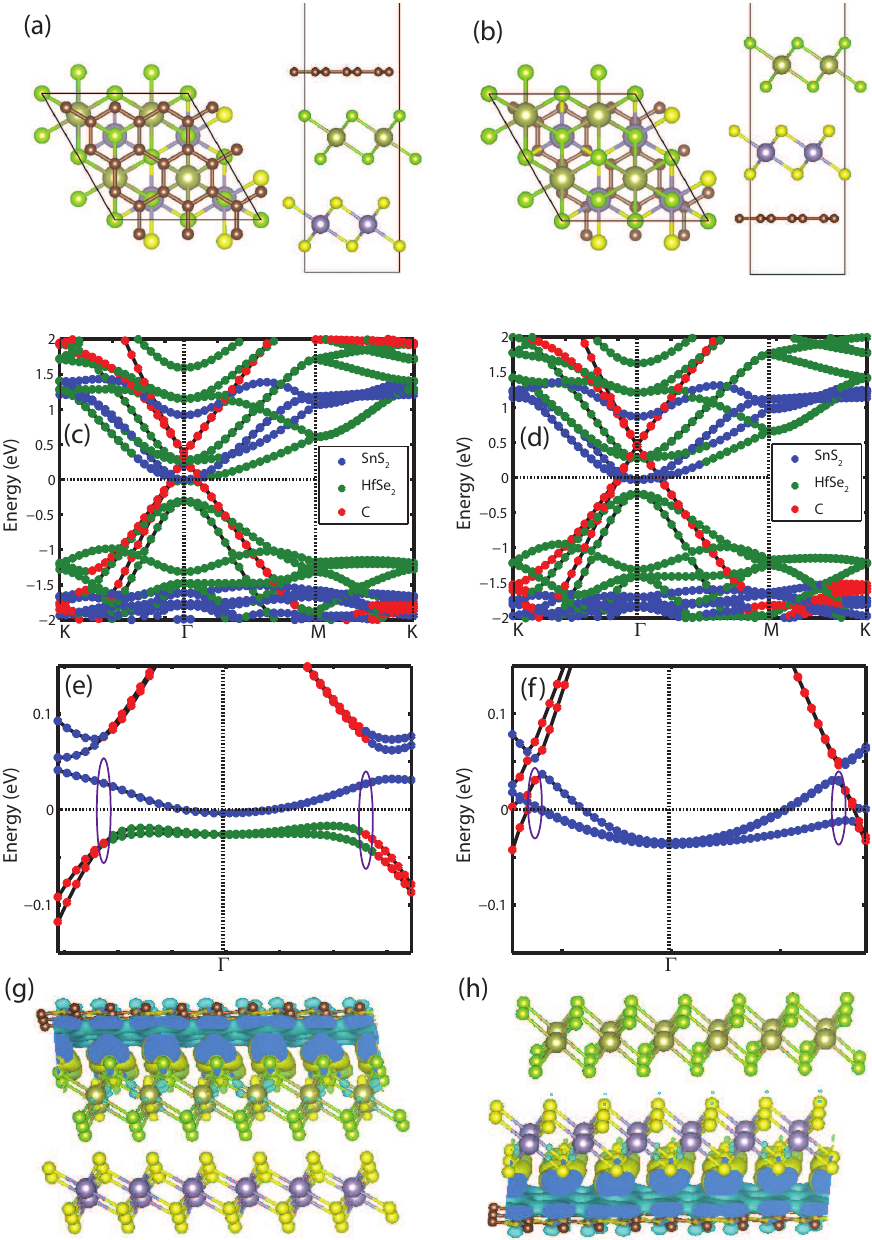}
\caption{\label{fig:AB_graphene_trilayer} 
Trilayer of graphene and AB stacked HfSe$_2$/SnS$_2$.
(a) Atomic structure of graphene on the HfSe$_2$ layer. 
(c) and (e) are the corresponding electronic structure plots. 
(e) focuses on the small energy range near the Fermi level.
(g) shows the charge transfer at the interface.
(b) Atomic structure of graphene on the SnS$_2$ layer.
(d) and (f) are the corresponding electronic structure plots. 
(f) focuses on the small energy range near the Fermi level.
(h) shows the charge transfer at the interface.
In (g) and (h), the charge accumulation and depletion is denoted by the yellow and blue color, respectively. 
The Fermi level is at $E=0$.
The purple circles in (e) and (f) indicate the anti-crossing of the
graphene hole band and the HfSe$_2$/SnS$_2$ conduction band.
}
\end{figure}

Placing graphene on either the HfSe$_2$ layer or the SnS$_2$ layer of the 
AA or AB stacked heterostructures results in charge transfer from the graphene
to the HfSe$_2$/SnS$_2$ heterostructure such that the 
graphene becomes p-type, the HfSe$_2$/SnS$_2$ becomes n-type,
and the Fermi level aligns above the CBM of the 
HfSe$_2$/SnS$_2$.
The structures and 
energy-momentum relations are shown in Figs. \ref{fig:AA_graphene_trilayer}(a-d) and \ref{fig:AB_graphene_trilayer}(a-d)
for graphene on the top or bottom of the AA or AB heterostructures, respectively.
%
%
The charge transfer $n_s$ can be estimated by position of the Dirac point of the graphene $E_D$ with
respect to the Fermi level $E_F$ as $n_s = \frac{1}{\pi (\hbar v)^2} (E_D - E_F)^2$.
With graphene on HfSe$_2$, $E_D - E_F = 0.335$ eV, and $n_s = 1.26 \times 10^{13}$ cm$^{-2}$.  
With graphene on SnS$_2$, $E_D - E_F = 0.430$ eV, and $n_s = 2.07 \times 10^{13}$ cm$^{-2}$.  
The calculated results are very close to the Bader charge transfer \cite{henkelman2006fast} results, $1.13 \times 10^{13}$ cm$^{-2}$ with graphene on HfSe$_2$ and $1.33 \times 10^{13}$ cm$^{-2}$ with graphene on SnS$_2$, obtained from VASP.
By electronic device standards, this sheet charge density transferred 
from the graphene into the HfSe$_2$/SnS$_2$ is large.
%
The electron transfer from the graphene to the  HfSe$_2$/SnS$_2$ 
is accompanied by a lowering of the potential of the layer
in contact with the graphene.
The region of the electronic bands around the Fermi level near $\Gamma$ is shown in 
Figs. \ref{fig:AA_graphene_trilayer}(e-f) and \ref{fig:AB_graphene_trilayer}(e-f).
In all cases, the lowest conduction band wavefunction is weighted more heavily towards the layer in contact
with the graphene.
This results in a negative Schottky barrier between the graphene and the conduction band of HfSe$_2$/SnS$_2$
for contact to either side of the heterostructure.

For a good contact, energy level alignment is critical, 
but there should also be 
coupling between the graphene and the HfSe$_2$/SnS$_2$ layers for electrons
to transfer easily between the two layers.
This coupling or interaction appears in the energy-momentum plots as an anti-crossing of the 
graphene and HfSe$_2$/SnS$_2$ bands.
The anti-crossing of the graphene band and the HfSe$_2$/SnS$_2$ conduction band
is shown in Figs. \ref{fig:AA_graphene_trilayer}(e,f) and \ref{fig:AB_graphene_trilayer}(e,f).
The color coding of the bands is the same as in 
Figs. \ref{fig:AA_graphene_trilayer}(c,d) and \ref{fig:AB_graphene_trilayer}(c,d).
In this commensurate Brillouin zone, the conduction band of the HfSe$_2$/SnS$_2$
is 3-fold degenerate (excluding spin), since the 6 $M$ points of the original 
HfSe$_2$/SnS$_2$ Brillouin zone are folded to $\Gamma$.
The Dirac cone of the graphene is two-fold degenerate, since the $K$ and $K'$ points
of the original graphene Brillouin zone are folded to $\Gamma$.
Where the bands anti-cross shown in the region of the vertical ellipses,
the interaction between the graphene and the HfSe$_2$ breaks the degeneracy,
so that the 2 Dirac cones and 3 conduction bands from the HfSe$_2$/SnS$_2$ can be clearly seen
in Figs. \ref{fig:AA_graphene_trilayer}(e) and \ref{fig:AB_graphene_trilayer}(e).
In Figs. \ref{fig:AA_graphene_trilayer}(e) and \ref{fig:AB_graphene_trilayer}(e),
the Dirac cone of the graphene anti-crosses with the conduction bands of the 
HfSe$_2$/SnS$_2$ with an energy splitting $\Delta$ on the order of 100 meV. 
A value for the coupling $t$ can be estimated from the energy splitting $\Delta$ of the splitting of bands at the crossing points.
Setting $\Delta = 2t$, gives $t \approx 50$ meV.
In the energy-momentum region where the bands anti-cross indicated by the vertical
ellipses, the wavefunction composition is a mix of orbitals from the graphene and the HfSe$_2$/SnS$_2$. 
The composition of
the three bands circled by the left ellipse in Fig. \ref{fig:AA_graphene_trilayer}(e) 
are, from lowest energy to highest energy:
(1) 26\% SnS$_2$, 44\% HfSe$_2$, and 29\% of graphene; (2) 22\% SnS$_2$, 38\% HfSe$_2$, and 40\% graphene; 
and (3) 48\% SnS$_2$, 47\% HfSe$_2$, and 4\% graphene.
The orbital composition of the bands circled by the right ellipse along the line from $\Gamma$
to $M$ are very similar.
Thus, both the energetic splitting and the orbital mixing indicate that there is significant 
coupling between the graphene and the HfSe$_2$ layer that should allow easy charge transfer
between the layers under applied bias.

The interaction of graphene with HfSe$_2$ is larger than with SnS$_2$,
and this is consistent with the orbital composition of the conduction bands of
HfSe$_2$ and SnS$_2$.
The conduction band of HfSe$_2$ has large Hf $d_{z^2}$ and Se $p_z$ 
components that would be expected to couple well to the C $p_z$ orbitals of the graphene.
The conduction band of SnS$_2$ has large Sn $s$ and S $p_x, p_y$ components.
The in-plane S $p_x, p_y$ orbitals would be expected to couple poorly to the C $p_z$ orbitals of the graphene.
When the graphene is placed on the SnS$_2$ layer, the bands near the Fermi level shown in
Fig. \ref{fig:AA_graphene_trilayer}(f) look qualitatively different
compared to the bands with graphene on the HfSe$_2$ layer.
At $\Gamma$, the 3 conduction bands of the HfSe$_2$/SnS$_2$ remain degenerate. 
All 3 of the conduction bands now lie 0.03 eV below the Fermi level, so that the 
Schottky barrier becomes more negative.
This is consistent with the fact that, as shown in Fig. \ref{fig:band_align_sep}, 
the conduction band of the SnS$_2$ is energetically
lower than that of HfSe$_2$, and the conduction band wavefunction
of the isolated HfSe$_2$/SnS$_2$ heterostructure
is more heavily weighted towards the SnS$_2$ layer as shown in Fig. \ref{fig:fig1}.
The energy alignment is more favorable for electrical contact, 
however the coupling between the graphene 
and the SnS$_2$ is considerably weaker. 
%
%
Now, the maximum energy splitting is $\approx 10$ meV giving an estimate for the
coupling of $t \approx 5$ meV.

The difference in coupling can also be seen in the composition of the conduction band wavefunctions
at the $\Gamma$ point.
For graphene on HfSe$_2$, at the $\Gamma$ point, the compositions of the three conduction bands nearest the Fermi energy,
from lowest to highest energy are:
(1) 40\% SnS$_2$, 54\% HfSe$_2$, and 5\% graphene; (2) 40\% SnS$_2$, 55\% HfSe$_2$, and 5\% of graphene; 
and (3) 48\% of SnS$_2$, 52\% of HfSe$_2$, and 0\% graphene. 
The highest split-off conduction band has its weight shifted more towards the SnS$_2$ layer compared to the
lower two conduction bands, and it has no graphene contribution.
With graphene on the SnS$_2$, the compositions of the three conduction bands nearest the Fermi energy
at the $\Gamma$ point, 
are all the same, and they are 65\% SnS$_2$, 35\% HfSe$_2$, and 0\% graphene.

The trends for graphene on the AB stacked structure are qualitatively the same as for graphene
on the AA structure.
As discussed with respect to Fig. \ref{fig:band_align_sep}, the interlayer coupling between the SnS$_2$ and HfSe$_2$
is weaker in the AB stacking arrangement compared to that with AA stacking.
Therefore, the wavefunction of the conduction band edge is more heavily weighted towards the 
SnS$_2$ in the isolated heterostructure.
In the AB structure, placing the graphene on the HfSe$_2$, reverses the weight of the bottom two conduction
bands in Fig. \ref{fig:AB_graphene_trilayer}, so that their compositions become
(1) 43\% SnS$_2$, 51\% HfSe$_2$, and 5\% graphene; and (2) 43\% SnS$_2$, 52\% HfSe$_2$, and 5\% graphene.
The spectral weight of the highest band is 52\% SnS$_2$, 48\% HfSe$_2$, and 0\% graphene.
The only qualitative difference between this structure and the AA structure is the slight shift in
orbital weight of the conduction band wavefunction towards the SnS$_2$.

The values of the interlayer couplings $t$ can be used in a tunneling Hamiltonian expression
to estimate the interlayer conductance
between the graphene and the HfSe$_2$/SnS$_2$ heterostructure 
when graphene is placed on either the SnS$_2$ layer or the HfSe$_2$ layer.
The interlayer conductivity can be written as,
\begin{equation}
\sigma_{c} = \frac{g_s g_G g_H e^2}{ {\cal{A}} h } \sum_{\kv}\int dE 
A_G(\kv; E) A_H(\kv; E) |t|^2 \frac{-\partial f(E-E_f)}{\partial E},
\label{eq:HT}
\end{equation}
where $A_G(\kv; E)$ and $A_H(\kv; E)$ are spectral functions of the 
graphene layer and the semiconductor heterostructure, respectively,
$g_s=2$ is the spin degeneracy, 
$g_G = 2$ is the graphene band degeneracy,
$g_H = 3$ is the HfSe$_2$/SnS$_2$ band degeneracy,
${\cal{A}}$ is the cross-sectional area,
$f(E-E_f)$ is the Fermi-Dirac factor,
$E_f$ is the Fermi level,
and $t$ is the interlayer coupling.
The spectral functions are 
$A_G(\kv; E) = \frac{\gamma}{(E+\hbar v_F k - \ep_D)^2 + \gamma^2/4}$, $A_H(\kv; E) = \frac{\gamma}{(E - \frac{\hbar^2 k^2}{2 m^*} - \ep_H)^2 + \gamma^2/4}$,
where
$\gamma$ is the lifetime broadening,
$m^* = 0.4 m_0$ is the effective mass obtained from the DFT bandstructures in Fig. \ref{fig:AA_graphene_trilayer}(c)-(d),
$v_F = 0.81 \times 10^6$ m/s is the Fermi velocity of graphene,
$\ep_D$ is the energy of the Dirac point,
and
$\ep_H$ is the energy of the conduction band minimum of the HfSe$_2$/SnS$_2$ heterostructure.
For a given transverse $\bf k$, the quantity
$T(E,\kv) = A_G(\kv; E) A_H(\kv; E) |t|^2$ is the transmission coefficient, and, as such,
its value must lie between 0 and 1. \cite{KZhou_rotated_MoS2}
The values of $\ep_D$ and $\ep_H$ are chosen such that the lower Dirac cone of the graphene
and the parabolic conduction band of the HfSe$_2$/SnS$_2$ intersect at the Fermi wavevector $k_F$,
$(E_F+\hbar v_F k_F - \ep_D) =  (E_F - \frac{\hbar^2 k_F^2}{2 m^*} - \ep_H) = 0$,
giving a maximum value for $T(E,\kv)$ of $16 |t|^2 / \gamma^2 \leq 1$.  
This, sets a lower limit on on the value for $\gamma$ of $\gamma \geq 4 |t|$.  
For graphene on HfSe$_2$, $t$ is large, $\approx 50$ meV, which pushes the limit of validity of the tunneling
Hamiltonian expression (\ref{eq:HT}), and it requires a large value for $\gamma$ of 200 meV.
The contact resistance is $R_C = 1/\sigma_{c}$,
and the resulting value for the contact resistance of graphene on HfSe$_2$ 
as shown in Fig. \ref{fig:AA_graphene_trilayer}(a) is 1 ${\rm m}\Omega \cdot \mu{\rm m}^2$.
With graphene on SnS$_2$, $t \approx 5$ meV, and the contact resistance is 
100 ${\rm m}\Omega \cdot \mu{\rm m}^2$.
The resistances scale as $|t|^2$, which accounts for the factor of 100 difference
in the contact resistances. 
Decreasing the value of $\gamma$ monotonically decreases $R_C$ by approximately
a factor of 5 as $\gamma$ is decreased from 200 meV to 20 meV.
These resistance values should be viewed as order-of-magnitude estimates. 
Both of these values are excellent in terms of the state-of-the art contact resistance
to 2D materials \cite{MoS2_nmat_phase_engg_contact},
and the lowest value is competitive with the best that has been achieved
in the well-developed semiconductors such as InGaAs \cite{Rc_InGaAs_Mohney_JAP13}.

\section{Summary and Conclusions}
\label{sec:conclusion}
Monolayer HfSe$_2$ and SnS$_2$ are closely lattice matched with a strain of less than 1\%,
an average lattice constant of 3.70 {\AA}, and bandgaps of 1.1 eV\cite{gaiser2004band} and 2.4 eV\cite{koda2016coincidence}, respectively. 
When the two materials are well-separated, but with a common Fermi level,
the HfSe$_2$ conduction band is 0.25 eV above the SnS$_2$ conduction band,
and the valence band of the HfSe$_2$ is more than 1 eV above the valence band of the SnS$_2$.
Such a band lineup in traditional three dimensional semiconductors leads
to a type II heterostructure in which the conduction band is on
one layer and the valence band is on the other. 
However, when the HfSe$_2$ and the SnS$_2$ are brought together to form a heterostructure,
the conduction band minimum at $M$ becomes a coherent superposition of the 
of the conduction band wavefunctions of the individual layers.
The conduction band wavefunction
is weighted 60\% on the SnS$_2$ and 40\% on the HfSe$_2$ for AA stacking
and 67\% on the SnS$_2$ and 33\% on the HfSe$_2$ for AB stacking.
There is no energy barrier for an electron to move between the two layers, since the conduction
band wavefunction is a coherent superposition of the orbitals of both layers.
A vertical electric field of 0.2 V/{\AA} pointing from the HfSe$_2$ layer to the SnS$2$ layer
reverses the weights of the conduction band wavefunction to approximately 70\% HfSe$_2$
and 30\% SnS$_2$.
In the SnS$_2$, the primary orbital contributions come from the $s$-orbital of the Sn
and the $p_{x,y}$ orbitals of the S.
In the HfSe$_2$, the primary orbital contributions come from the $d_{z^2}$ orbital of the Hf
and the $p_z$ orbital of the Se.
The valence band maximum at $\Gamma$ is localized on the HfSe$_2$ layer, and its dominant 
orbital contributions come from the $p_x$ and $p_y$ orbitals of the Se.
The calculated HSE bandgap of the AA and AB heterostructures are 0.88 eV and 0.89 eV, respectively.

A $3 \times 3$ supercell of graphene is almost perfectly lattice matched to a $2 \times 2$ supercell
of HfSe$_2$/SnS$_2$ with a lattice mismatch of 0.1\%. 
The trilayer heterostructure is stable with negative formation energies, and the formation
energy with graphene on the HfSe$_2$ is approximately 50 meV more negative than with graphene
on the SnS$_2$. 
This indicates a stronger interaction of the graphene with the HfSe$_2$, which is consistent with
the results from the electronic structure calculations. 
A charge density on the order of $10^{13} / {\rm cm}^2$ transfers from the graphene to the HfSe$_2$/SnS$_2$
resulting in a Fermi level that aligns within the conduction band of the the HfSe$_2$/SnS$_2$
and a negative Schottky barrier contact for electron injection into the conduction band.
The coupling of the graphene to the HfSe$_2$ is approximately 10 times larger than the coupling
of the graphene to the SnS$_2$, and this is consistent with the $d_{z^2}$ and $p_z$ orbital 
composition of the  HfSe$_2$ conduction band compared to the $s$ and $p_{x,y}$ composition of the
SnS$_2$ conduction band. 
A tunneling Hamiltonian estimate for the contact resistance of graphene on the HfSe$_2$
layer
versus graphene on the SnS$_2$ layer gives contact resistances of 1 m$\Omega \mu {\rm m}^2$
and 100 m$\Omega \mu {\rm m}^2$, respectively. 
Both values would be considered exceptional for 2D materials, and the lowest value is 
competitive with lowest contact resistances measured in 3D semiconductors such as InGaAs.


\vspace{0.5in}
\noindent{\em Acknowledgements:}
This work is supported in part by FAME, one of six centers of STARnet, 
a Semiconductor Research Corporation program sponsored by MARCO and DARPA,
and by the NSF EFRI-143395. 
This work used the Extreme Science and Engineering Discovery Environment (XSEDE), 
which is supported by National Science Foundation grant number ACI-1053575.

%

\end{document}